\documentclass[a4paper,11pt]{article}
\usepackage{pos}

\usepackage{graphicx}
\usepackage[makeroom]{cancel}
\usepackage{subcaption}

\usepackage{xcolor}
\colorlet{alert}{red!60!black}
\colorlet{example}{green!60!black}
\colorlet{structure}{blue!60!black}

\newcommand{\mnote}[1]{}                   

\def\lsim{\mathrel{\rlap{\lower4pt\hbox{\hskip1pt$\sim$}}
    \raise1pt\hbox{$<$}}}                
\def\gsim{\mathrel{\rlap{\lower4pt\hbox{\hskip1pt$\sim$}}
    \raise1pt\hbox{$>$}}}                

\newcommand{\half}{\mbox{\small $\frac{1}{2}$}}          

\begin{document}


\title{
\vspace*{-0.75cm}
\begin{minipage}{\textwidth}
\begin{flushright}
\texttt{\footnotesize
PoS(LATTICE2022)412  \\
ADP-23-04/T1214      \\
DESY-23-017          \\
Liverpool LTH 1331   \\
}
\end{flushright}
\end{minipage}\\[15pt]
\vspace*{+0.75cm}
        Quasi-degenerate baryon energy states, the Feynman--Hellmann 
        theorem and transition matrix elements}

\ShortTitle{Quasi-degenerate baryon energy states \ldots}

\author*[a,1]{M.~Batelaan}
\author[a]{K.~U.~Can}
\author*[b,1]{R.~Horsley}
\author[c]{Y.~Nakamura}
\author[d]{H.~Perlt}
\author[e]{P.~E.~L.~Rakow}
\author[f]{G.~Schierholz}
\author[g]{H.~St\"uben}
\author[a]{R.~D.~Young}
\author[a]{J.~M.~Zanotti}

\affiliation[a]{CSSM, Department of Physics,
               University of Adelaide, Adelaide SA 5005, Australia}
\affiliation[b]{School of Physics and Astronomy, University of Edinburgh,
                Edinburgh  EH9 3FD, UK}
\affiliation[c]{RIKEN Center for Computational Science,
                Kobe, Hyogo 650-0047, Japan}
\affiliation[d]{Institut f\"ur Theoretische Physik,
                Universit\"at Leipzig, 04109 Leipzig, Germany}
\affiliation[e]{Theoretical Physics Division,
                Department of Mathematical Sciences,
                University of Liverpool, \\
                Liverpool L69 3BX, UK}
\affiliation[f]{Deutsches Elektronen-Synchrotron DESY,
                Notkestr. 85, 22607 Hamburg, Germany}
\affiliation[g]{Universit\"at Hamburg, Regionales Rechenzentrum,
                20146 Hamburg, Germany}

\note{For the QCDSF-UKQCD-CSSM Collaborations}

\emailAdd{mischa.batelaan@adelaide.edu.au, rhorsley@ph.ed.ac.uk}

\abstract{The standard method for determining matrix elements in lattice QCD 
          requires the computation of three-point correlation functions. 
          This has the disadvantage of requiring two large time separations:
          one between the hadron source and operator and the other from 
          the operator to the hadron sink. Here we consider an alternative 
          formalism, based on the Dyson expansion leading to the 
          Feynman-Hellmann theorem, which only requires the computation 
          of two-point correlation functions. Both the cases of degenerate 
          energy levels and quasi-degenerate energy levels which correspond 
          to diagonal and transition matrix elements respectively can be
          considered in this formalism. As an example numerical results 
          for the Sigma to Nucleon vector transition matrix element 
          are presented.}

\FullConference{%
  The 39th International Symposium on Lattice Field Theory (Lattice2022),\\
  8-13 August, 2022 \\
  Bonn, Germany 
}

\maketitle


\section{Introduction}


A major pursuit of lattice QCD is the determination of non-perturbative
matrix elements generically given by 
$\langle H^\prime | \hat{\cal O} | H \rangle$ where $H$ is a hadron such 
as $H \sim \bar{q}q$ (meson) or $H \sim qqq$ (baryon) and the operator 
$\hat{\cal O} \sim \bar{q}\gamma q \sim J$ or $\hat{\cal O} \sim FF$ 
or even more complicated $\hat{\cal O} \sim JJ$. While the usual 
approach is to compute ratios of $3$-point correlation functions to 
$2$-point correlation functions in these talks we will describe an 
alternative method based on the Feynman--Hellmann theorem, which only 
involves computing perturbed $2$-point correlation functions. 
In \cite{QCDSF:2017ssq} we discussed this for nucleon scattering. 
However this required degenerate energy states. We shall now
describe a generalisation of the Feynman--Hellmann approach from 
the determination of nucleon matrix elements with degenerate energy states to 
near-degenerate or `quasi-degenerate' energy states, \cite{qcdsf_fh}.

In these talks we shall first discuss the theory behind the 
Feynman--Hellmann approach via the transfer matrix to a computation 
of $2$-pt correlation functions with particular application to 
quasi-degenerate states. We employ the Dyson expansion to reduce the 
problem to a Generalised EigenVector Problem (GEVP) giving avoided
energy levels. As examples we first consider $N$ scattering for flavour 
diagonal matrix elements. However naturally our approach is valid 
for decay or transition matrix elements, for example the $\Sigma \to N$ 
transition. (These matrix elements occur in semi-leptonic hyperon
decays and provide an alternative approach to determining the CKM
matrix element $V_{us}$, \cite{Cabibbo:2003ea}.)
In both cases we give sketches of avoided energy levels.
We then turn to a numerical simulation for the vector current for
this transition matrix element, confirming our previous theoretical
discussion. Finally we give our conclusions. For more details, 
see \cite{qcdsf_fh}.


\section{The Feynman--Hellmann approach}


In this section we shall give some mathematical details of our
Feynman--Hellmann (FH) approach. We employ the Hamiltonian formalism and
regard Euclidean time (at least) as continuous. Although our approach
is to consider the the $2$-point nucleon correlation function, it is valid
for all hadrons. We shall make some comments about the introduction 
of spin later. We have
\begin{eqnarray}
   C_{\lambda\,B^\prime B}(t)
      = {}_\lambda\langle 0| 
          \hat{\tilde{B}}^\prime(0;\vec{p}^\prime) \hat{S}(\vec{q})^t 
               \hat{\bar{B}}(0,\vec{0})|0\rangle_\lambda \,,
\label{2pt_cf_original}
\end{eqnarray}
where the source $\hat{\bar{B}}(0,\vec{0})$ is spatial (for simplicity
placed at the origin $\vec{0}$) and contains all momenta, while the
sink $\tilde{B}^\prime(0;\vec{p}^\prime)$ picks out a particular momentum
$\vec{p}^\prime$. $\hat{S}$ is the $\vec{q}$-dependent transfer matrix 
$\hat{S}(\vec{q}) = e^{-\hat{H}(\vec{q})}$ in the presence of a perturbed 
Hamiltonian
\begin{eqnarray}
   \hat{H}(\vec{q}) = \hat{H}_0 
        - \sum_\alpha \lambda_\alpha \hat{\tilde{{\cal O}}}_\alpha(\vec{q}) \,,
\end{eqnarray}
with
\begin{eqnarray}
   \hat{\tilde{{\cal O}}}(\vec{q})
      = \int_{\vec{x}} \left( \hat{O}(\vec{x})e^{i\vec{q}\cdot\vec{x}}
                            + \hat{O}^\dagger(\vec{x}) e^{-i\vec{q}\cdot\vec{x}}
                     \right) \,.
\end{eqnarray}
In the large box-size limit, we pick out the ground state of the 
perturbed Hamiltonian, $|0\rangle_\lambda$ as indicated in 
eq.~(\ref{2pt_cf_original}). At leading order (considered here)
we can drop the $\alpha$ index. (At higher orders this is not possible.)
Also as we write $\lambda_\alpha = |\lambda_\alpha|\zeta_\alpha$ 
($\zeta_\alpha = \pm 1, \pm i$) then any phase can be absorbed into $\hat{O}$ 
and we can consider positive $\lambda_\alpha$ only. 

\begin{figure}[htb]
   \begin{center}
      \includegraphics[width=3.50cm]{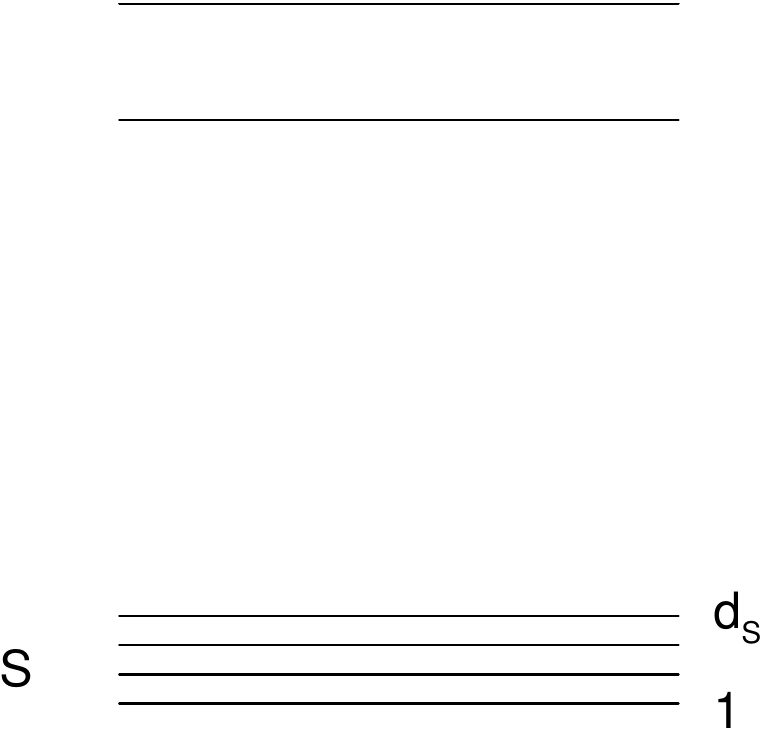}
   \end{center}
   \caption{A sketch of the energy levels. The set of 
            quasi-degenerate energy states are denoted by $S$, labelled
            from $1$ to $d_S$. These states are well separated from other
            higher states.}
\label{sketch_energy_levels}
\end{figure}

We consider the physical situation with quasi-degenerate 
energies as shown in Fig.~\ref{sketch_energy_levels},
taking the $d_S$ quasi-degenerate states to be well separated from any
higher energy states as shown in the figure. Their energies are defined by
\begin{eqnarray}
   \hat{H}_0 |B_r(\vec{p}_r) \rangle 
       = E_{B_r}(\vec{p}_r) |B_r(\vec{p}_r) \rangle \,,
               \quad r = 1\,, \ldots\,, d_S \,,
\end{eqnarray}
where
\begin{eqnarray}
   E_{B_r}(\vec{p}_r) = \bar{E} + \epsilon_r \,,
\end{eqnarray}
$\bar{E}$ being some typical quasi-degenerate energy (for example their 
average energy). They are well separated from higher energy states:
\begin{eqnarray}
   \hat{H}_0 |X(\vec{p}_X) \rangle 
      = E_X(\vec{p}_X) |X(\vec{p}_X) \rangle \quad E_X \gg \bar{E} \,.
\end{eqnarray}
Practically we thus take the quasi-degenerate states as the lowest energy 
states.

For the matrix elements corresponding to the quasi-degenerate energy levels
in Fig.~\ref{sketch_energy_levels} we have a relation between the various 
momenta. Using $\hat{O}(\vec{x}) 
                 = e^{-i\hat{\vec{p}}\cdot\vec{x}}\,\hat{O}(\vec{0}) \,
                                      e^{i\hat{\vec{p}}\cdot\vec{x}}$
we soon see that
\begin{eqnarray}
   \langle B(\vec{p}_r)| \hat{\tilde{{\cal O}}}(\vec{q}) 
                                     | B(\vec{p}_s) \rangle
      = \langle B_r(\vec{p}_r)| \hat{O}(\vec{0})
                                     | B_s(\vec{p}_s) \rangle
                \, \delta_{\vec{p}_r,\vec{p}_s+\vec{q}}
        + \langle B(\vec{p}_r)| \hat{O}^\dagger(\vec{0})
                                       |B(\vec{p}_s) \rangle
                \, \delta_{\vec{p}_r,\vec{p}_s-\vec{q}} \,.
\end{eqnarray}
So matrix elements step up or down in $\vec{q} \not= \vec{0}$
\begin{eqnarray}
   \vec{p}_r = \vec{p}_s + \vec{q}\,, \quad \mbox{or} \quad
   \vec{p}_r = \vec{p}_s - \vec{q}\,, 
\end{eqnarray}
i.e.\ momentum conservation. 
(For $\vec{q}\to\vec{0}$ the states coalesce, a special case.)
We see immediately that diagonal matrix 
elements vanish. So (quasi)-degenerate states have to mix with one other
and we must consider degenerate perturbation theory.
We expect that each step up or down corresponds to another order in $\lambda$ 
as can be seen from the forthcoming Dyson expansion. So for example 
the $O(\lambda^2)$ term gives Compton-like amplitudes 
$\sim \langle\ldots|\hat{O}_\alpha \hat{O}_\beta|\ldots\rangle$.
In this case both step up and step down are now possible:
$\vec{p}_s \to \ \vec{p}_s\pm\vec{q} \to \vec{p}_s$ which are relevant 
for the forward Compton amplitude in e.g.\ DIS which is considered 
elsewhere, \cite{Chambers:2017dov,Can:2020sxc,Can:2022chd}. 

Now insert two complete sets of unperturbed states%
\footnote{We use the (lattice) normalisation $\langle X|X \rangle = 1$.
To convert to other normalisations use 
$|X\rangle \to |X\rangle / \sqrt{\langle X|X \rangle}$ and
$|0\rangle \to |0\rangle$. In particular for the standard relativistic 
normalisation we have $\langle X|X \rangle = 2E_X$ (used in 
eq.~(\ref{DeltaE_decay})).}
\begin{eqnarray}
   \sumint_{X(\vec{p_X})} |X(\vec{p}_X) \rangle \, \langle X(\vec{p}_X)|
   \equiv
   \sum_r \underbrace{|B_r(\vec{p}_r)\rangle\langle B_r(\vec{p}_r)|}_
                      {\rm of \, interest} 
      + \sumint_{E_X \gg \bar{E}} 
          \underbrace{|X(\vec{p}_X) \rangle \, \langle X(\vec{p}_X)|}_
                      {\rm higher \, states}
      = \hat{1} \,,
\label{spectral_decomp}
\end{eqnarray}
before and after $\hat{S}^t$ to give
\begin{eqnarray}
   C_{\lambda\,B^\prime B}(t)
     = \sumint_{X(\vec{p}_X)} \sumint_{Y(\vec{p}_Y)} 
            {}_\lambda\langle 0 |
                 \hat{\tilde{B}}^\prime(\vec{p}^\prime) 
                                              | X(\vec{p}_X) \rangle \,
             {\langle X(\vec{p}_X) | \hat{S}_\lambda(\vec{q})^t 
                               | Y(\vec{p}_Y) \rangle} \, 
       \langle Y(\vec{p}_Y) | \hat{\bar{B}}(\vec{0}) 
                                |0\rangle_\lambda \,.
\end{eqnarray}
Time dependent perturbation theory via the Dyson Series
iterates the operator identity
\begin{eqnarray}
   e^{-(\hat{H}_0 - \lambda_\alpha\hat{\tilde{{\cal O}}}_\alpha)t}
      &=& e^{-\hat{H}_0t} 
           + \lambda_\alpha\, \int_0^t dt^\prime\, e^{-\hat{H}_0(t - t^\prime)} \, 
                    \hat{\tilde{{\cal O}}}_\alpha \, 
                      e^{-(\hat{H}_0-
                       \xcancel{\lambda_\beta\hat{\tilde{{\cal O}}}_\beta}) 
                         t^\prime} \,,
\end{eqnarray}
where at leading order we simply drop the perturbation under the integral
as indicated. As mentioned before the $O(\lambda_\alpha\lambda_\beta)$ term 
would give Compton like amplitudes 
$\sim \langle\ldots|\hat{O}_\alpha \hat{O}_\beta|\ldots\rangle$.
Considering the possible pieces separately from 
eq.~(\ref{spectral_decomp})  gives finally the result, \cite{qcdsf_fh},
\begin{eqnarray}
   C_{\lambda\,B^\prime B}(t)
      = \sum_{i=1}^{d_S} w_{B^\prime}^{(i)} \bar{w}^{(i)}_B \, e^{-E^{(i)}_\lambda t}
           + \ldots \,,
\label{C_wwbar}
\end{eqnarray}
with perturbed energies
\begin{eqnarray}
   E_\lambda^{(i)} = \bar{E} - \mu^{(i)} \,, \quad i = 1, \ldots, d_S \,,
\end{eqnarray}
where $\mu^{(i)}$ are the eigenvalues%
\footnote{$D_{rs}$ is decomposed as 
$D_{rs} = \sum_{i=1}^{d_S} \mu^{(i)} e^{(i)}_r e^{(i)\,*}_s$.}
of the $d_S \times d_S$ Hermitian matrix $D_{rs}$ defined by
\begin{eqnarray}
   D_{rs} = - \epsilon_r \delta_{rs}
           + \lambda \langle B_r(\vec{p}_r) | 
                             \hat{\tilde{{\cal O}}}(\vec{q}) 
                                          | B_s(\vec{p}_s) \rangle \,.
\label{D_def}
\end{eqnarray}
Furthermore in eq.~(\ref{C_wwbar}) we have
\begin{eqnarray}
   w_{B^\prime}^{(i)} = \sum_{r=1}^{d_s} Z^{B^\prime}_r e_r^{(i)}\,, 
   \quad \mbox{and} \quad
   \bar{w}_{B}^{(i)} = \sum_{s=1}^{d_s} \bar{Z}^B_s e_s^{(i)*} \,,
\end{eqnarray}
where $\vec{e}^{(i)}$, $i = 1, \ldots, d_s$ are the $d_s$ eigenvectors of 
$D_{rs}$ and the wavefunctions are given by
\begin{eqnarray}
   Z^{B^\prime}_r = {}_\lambda\langle 0 | \hat{\tilde{B}}^\prime(\vec{p}^\prime)
                                    | B_r(\vec{p}_r) \rangle_\lambda \,, 
   \quad \mbox{and} \quad
   \bar{Z}^B_s = {}_\lambda\langle B_s(\vec{p}_s) | \hat{\bar{B}}(\vec{0}) 
                                                  |0 \rangle_\lambda \,,
\end{eqnarray}
where the states $|B_s(\vec{p}_s)\rangle_\lambda$ are defined by
\begin{eqnarray}
   |B_s(\vec{p}_s)\rangle_\lambda
      = |B_s(\vec{p}_s)\rangle
         + \lambda \sumint_{E_Y \gg \bar{E}}
                    |Y(\vec{p}_Y)\rangle  \,
                    { \langle Y(\vec{p}_Y)|\hat{\tilde{{\cal O}}}(\vec{q})
                      |B_s(\vec{p}_s)\rangle  
                    \over
                    E_Y - E_{B_s} } \,.
\label{state_lambda}
\end{eqnarray}
We see that there is a factorisation where the unwanted $|Y\rangle$ states 
have been absorbed into a time independent renormalisation of the 
wavefunction.

So from eq.~(\ref{C_wwbar}) we see that the problem is now `reduced' 
to a GEVP or Generalised EigenVector Problem,
\cite{Luscher:1990ck,Blossier:2009kd},  which can be applied to 
determine the energy eigenvalues $E_\lambda^{(i)}$ as described in
section~\ref{lattice}.

In principle this means that we can extend the computation to include
lower energy states $|Z\rangle$ in the spectrum, with $E_Z \ll \bar{E}$,
i.e.\ again well separated from the quasi-energy states. If there are
such states present in eq.~(\ref{spectral_decomp}) then we need to avoid 
any transitions between these states and either the quasi-degenerate
states or the higher energy states, as these will have a term $\sim e^{-E_Zt}$
and hence will be the leading term in eq.~(\ref{C_wwbar}). 
This can be achieved by a possible mixture of vanishing overlaps with 
these states, vanishing matrix elements and regarding them as extra terms 
in the GEVP. We do not consider this lower energy case further here.

Finally note that the above result is true for general source and 
sink operators. If we are able to set $\hat{B}^\prime$ and $\hat{B}$ `close'
to $\hat{B}_r$ and $\hat{B}_s$ respectively then the above 
expressions simplify and we have
\begin{eqnarray}
   w_r^{(i)} = Z_r e_r^{(i)}\,, \quad \mbox{and} \quad
   \bar{w}_s^{(i)} = \bar{Z}_s e_s^{(i)*} \,.
\label{w_simple}
\end{eqnarray}


\section{Examples}
\label{examples}


Let us consider a $d_S=2$-fold case: $r$, $s$ = $1$, $2$.
Then due to the step up or down in $\vec{q}$ for the matrix element
we must have
\begin{eqnarray}
   \langle B_r(\vec{p}_r) | \hat{\tilde{{\cal O}}}(\vec{q}) 
                         | B_s(\vec{p}_s) \rangle
      = \left( \begin{array}{cc}
                  0  & a^*  \\
                  a  & 0    \\
               \end{array}
         \right)_{rs} \,,
      \quad \mbox{where} \quad
   a = \langle B_2(\vec{p}_2)|\hat{O}(\vec{0})|B_1(\vec{p}_1)\rangle \,.
\label{matrix_a}
\end{eqnarray}
Diagonalising $D_{rs}(\vec{p},\vec{q})$ in eq.~(\ref{D_def}) gives
upon solving the quadratic equation the eigenvalues $\mu_{\pm}$
giving energies
\begin{eqnarray}
   E^{(\pm)}_\lambda
     = \bar{E} - \mu_\pm 
     = {1 \over 2}(E_2 + E_1) \mp {1 \over 2}\Delta E_\lambda \,,
\label{E_pm}
\end{eqnarray}
with
\begin{eqnarray}
   \Delta E_\lambda 
     = E^{(-)}_\lambda - E^{(+)}_\lambda
     = \sqrt{ (E_2 - E_1)^2
               + 4\lambda^2 |a|^2 } \,.
\label{DeltaE}
\end{eqnarray}
A flavour diagonal matrix element is given from nucleon scattering 
where we have
\begin{eqnarray}
   O(\vec{x}) \sim (\bar{u}\gamma u)(\vec{x}) - (\bar{d}\gamma d)(\vec{x})\,,
   \quad \mbox{and} \quad
   \underbrace{|B_1(\vec{p}_1)\rangle = |N(\vec{p})\rangle}_
              {E_{B_1}(\vec{p}_1) \equiv E_N(\vec{p}) = \bar{E} + \epsilon_1}\,,
   \quad
   \underbrace{|B_2(\vec{p}_2)\rangle = |N(\vec{p}+\vec{q})\rangle}_
              {E_{B_2}(\vec{p}_2) \equiv E_N(\vec{p}+\vec{q})
                                             = \bar{E} + \epsilon_2} \,.
\label{diagonal_case}
\end{eqnarray}
In general we have quasi-degenerate energy states, but it is easy to
choose $\vec{p}$ and $\vec{q}$ so the energies are degenerate
$E_N(\vec{p}+\vec{q}) = E_N(\vec{p})$, \cite{QCDSF:2017ssq}.
(A similar situation occurs if we consider $E_N(\vec{p}-\vec{q})$ instead.) 
For flavour transition matrix elements for example $\Sigma(sdd) \to N(udd)$ 
decay we have
\begin{eqnarray}
   O(\vec{x}) \sim (\bar{u} \gamma s)(\vec{x}) \,,
   \quad \mbox{and} \quad
   \underbrace{|B_1(\vec{p}_1)\rangle = |\Sigma(\vec{p})\rangle}_
              {E_{B_1}(\vec{p}_1) \equiv E_\Sigma(\vec{p}) 
                                 = \bar{E} + \epsilon_1}\,,
   \quad
   \underbrace{|B_2(\vec{p}_2)\rangle 
                                 = |N(\vec{p}+\vec{q})\rangle}_
              {E_{B_2}(\vec{p}_2) \equiv E_N(\vec{p}+\vec{q})
                                 = \bar{E} + \epsilon_2} \,.
\label{transition_me}
\end{eqnarray}
As $M_\Sigma \not= M_N$ then we now usually have quasi-degenerate energy states. 
Both cases (diagonal and transition matrix elements) thus have a similar 
structure.

\begin{figure}

\begin{minipage}{0.40\textwidth}

   \hspace*{0.25in}
   \includegraphics[width=5.00cm]{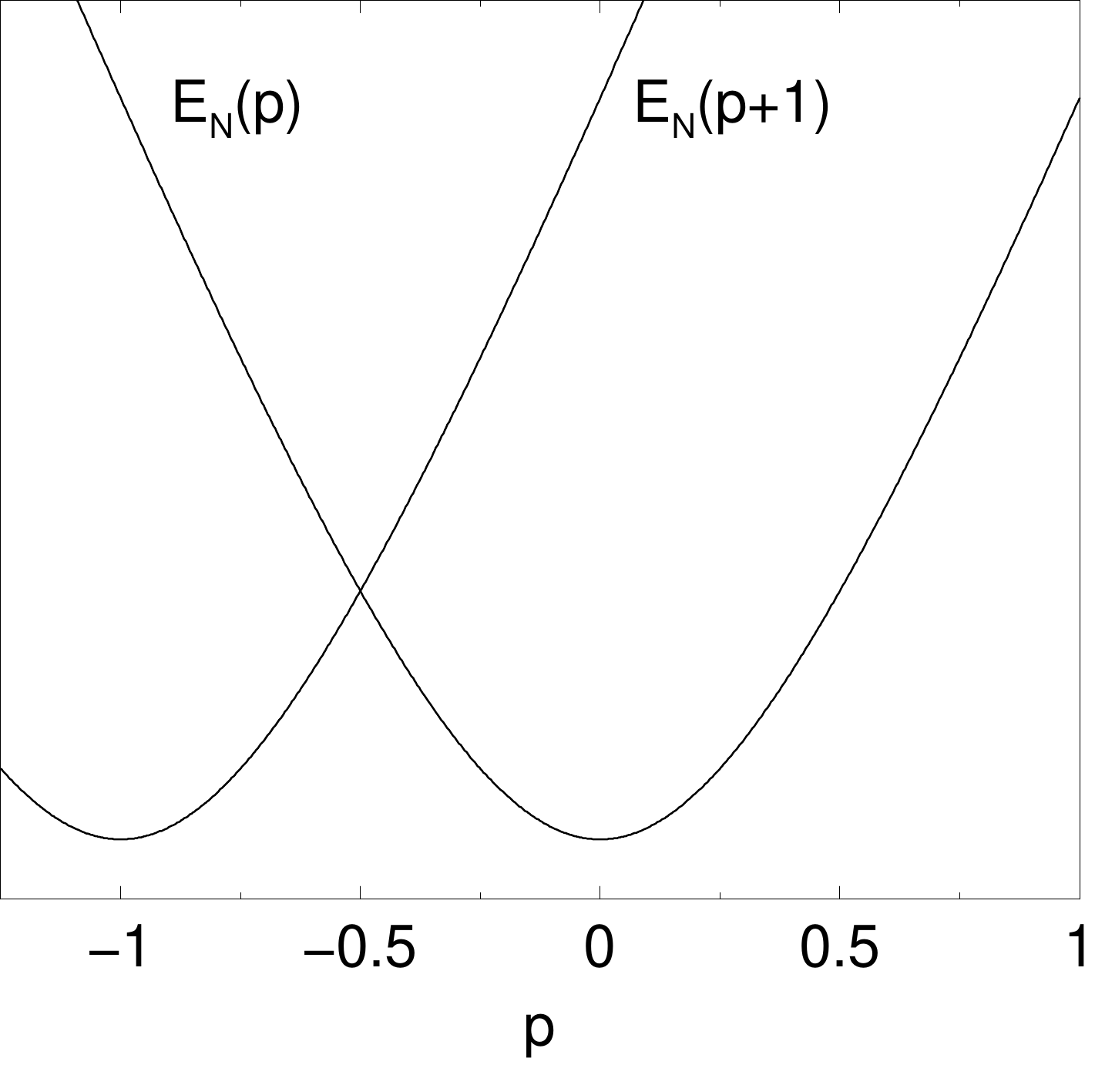}

\end{minipage}
\begin{minipage}{0.10\textwidth}
   \hspace*{0.15in}
   $\Rightarrow$
   \hspace*{-0.15in}
\end{minipage}
\begin{minipage}{0.40\textwidth}

   \includegraphics[width=5.00cm]{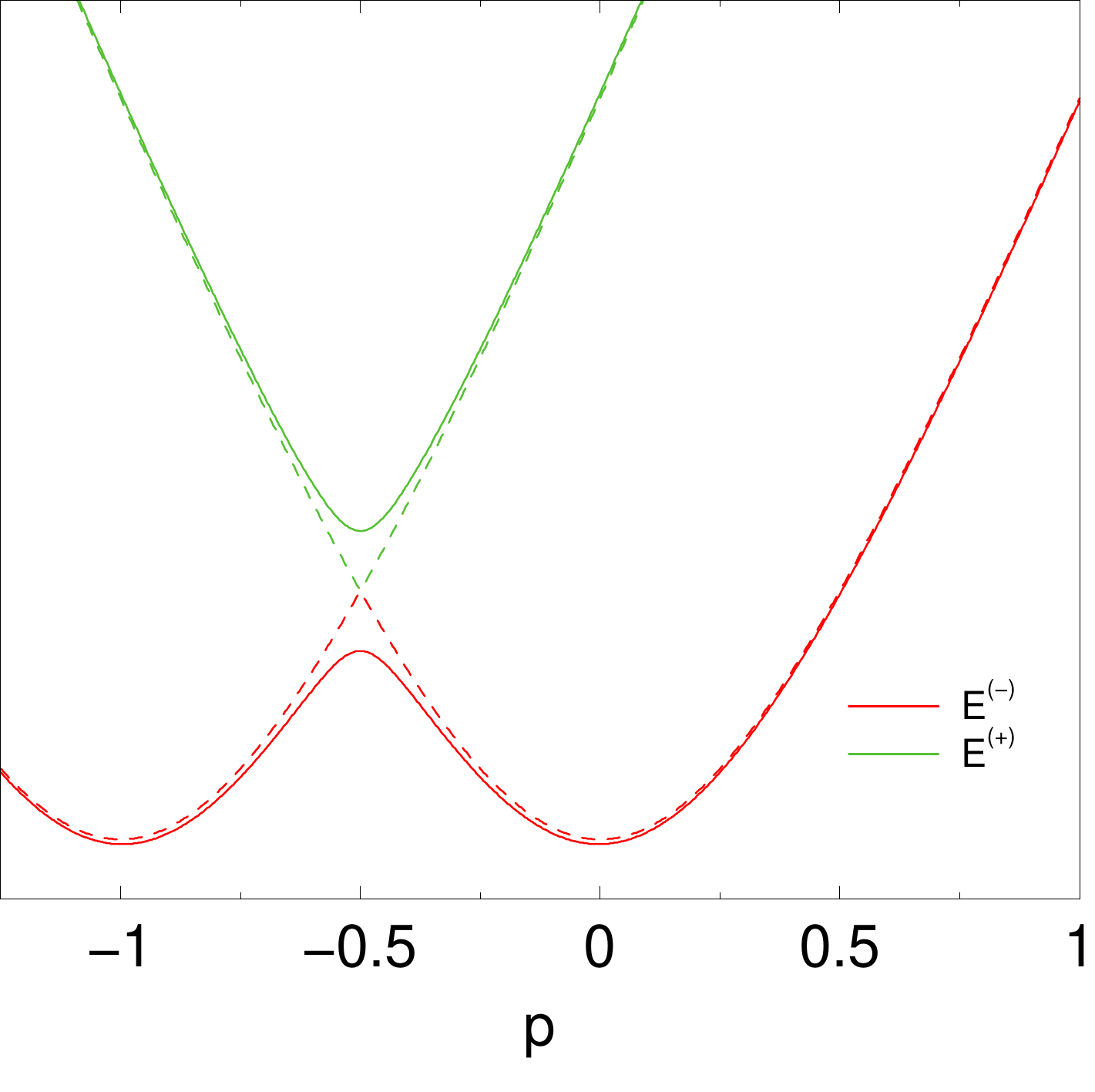}

\end{minipage}

\caption{Left panel: Plotting $p$ versus $E_N$ for the free case when 
         we have quasi-degenerate energies taking units where $q=1$ at 
         $p \approx -1/2$. 
         Right panel: The interacting case showing `avoided energy levels'.
         The $\lambda \to 0$ or free case is shown as dashed lines.}
\label{sketch_Nscat}
\end{figure}

\begin{figure}

\begin{minipage}{0.40\textwidth}

   \hspace*{0.25in}
   \includegraphics[width=5.00cm]{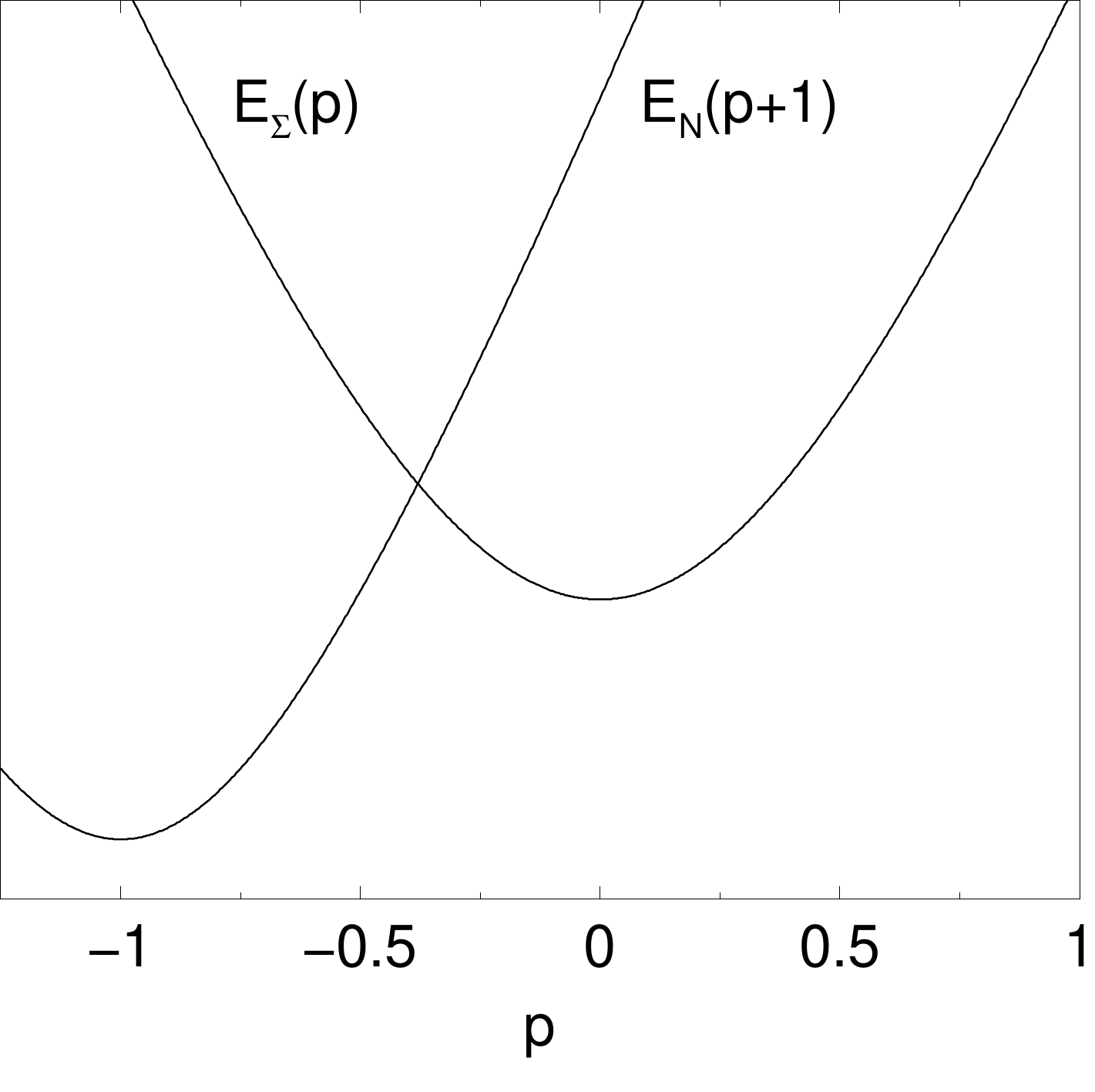}

\end{minipage}
\begin{minipage}{0.10\textwidth}
   \hspace*{0.15in}
   $\Rightarrow$
   \hspace*{-0.15in}
\end{minipage}
\begin{minipage}{0.40\textwidth}

   \includegraphics[width=5.00cm]{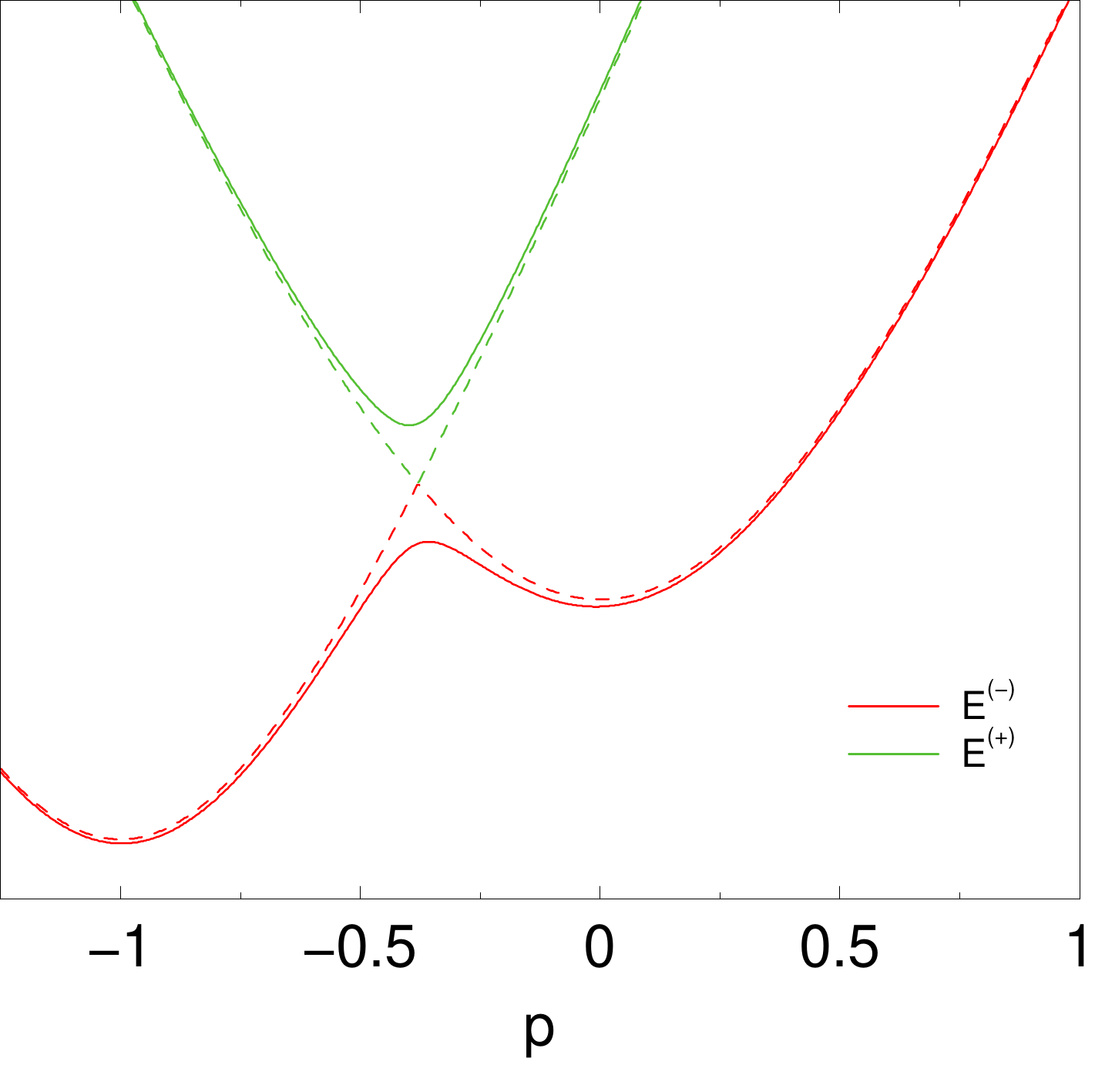}

\end{minipage}

\caption{Left panel: The free case when we have quasi-degenerate energies
         again taking units where $q=1$.
         Right panel: The interacting case showing `avoided energy levels'.}
\label{sketch_decay}
\end{figure}

We now illustrate these results with a series of (exaggerated) 
$1$-dimensional sketches. For nucleon scattering, eq.~(\ref{diagonal_case}),
we have the situation depicted in Fig.~\ref{sketch_Nscat}.
We focus on the degeneracy when $E_N(p) = E_N(p+q)$ at $p = -q/2$ where in 
addition $\epsilon_1 = 0 = \epsilon_2$. When the free case (left panel of
Fig.~\ref{sketch_Nscat}) becomes the interacting case (right panel of
Fig.~\ref{sketch_Nscat}) we have the phenomenon of `avoided energy level
crossing' when the energy levels do not cross. The sketch curves are based on 
previously derived formulae for $E^{(+)}$, $E^{(-)}$ in eq.~(\ref{E_pm}) 
and occur because in eq.~(\ref{DeltaE}) the square-root is always positive. 
Again note that a similar situation arises when $E_N(p) = E_N(p-q)$ at 
$p = q/2$ (not shown in the sketch). A similar situation occurs for 
$\Sigma \to N$ decay as illustrated in Fig.~\ref{sketch_decay}.
Again we have a degeneracy: $E_\Sigma(p) = E_N(p+q)$ which is now shifted
slightly to smaller $p$, as indicated in the figure.

\begin{figure}

\begin{minipage}{0.40\textwidth}

   \hspace*{0.25in}
   \includegraphics[width=4.70cm]{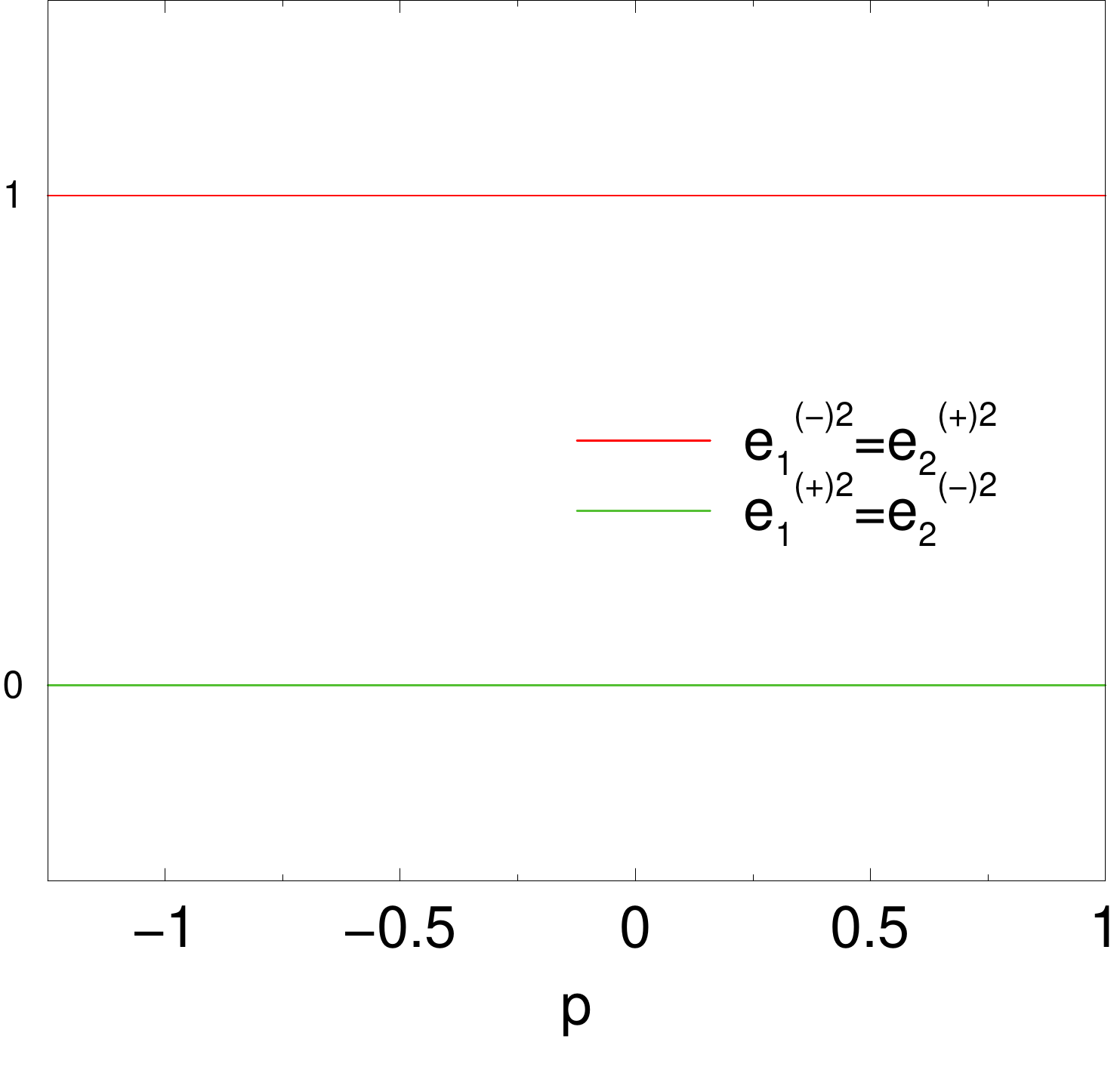}

\end{minipage}
\begin{minipage}{0.10\textwidth}
   \hspace*{0.15in}
   $\Rightarrow$
   \hspace*{-0.15in}
\end{minipage}
\begin{minipage}{0.40\textwidth}

   \includegraphics[width=4.70cm]{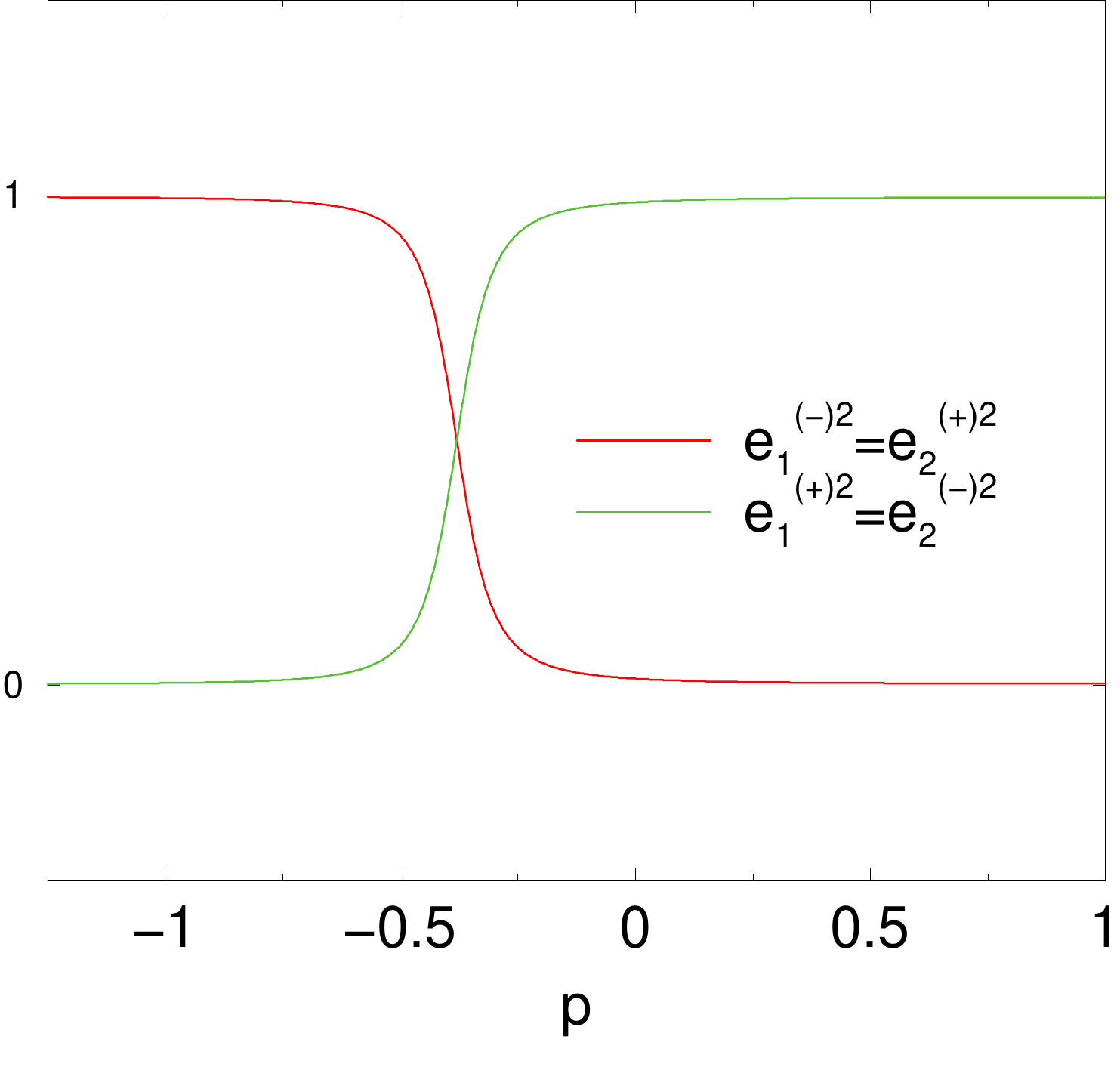}

\end{minipage}

\caption{Left panel: The free case where we have plotted $e_1^{(-)\,2}$
         and $e_2^{(-)\,2}$ against $p$, again taking units where $q=1$.
         Right panel: The interacting case showing the change of 
         state.}
\label{sketch_eigenvec_decay}
\end{figure}

The eigenvectors $e^{(\pm)}_r$ are given by
\begin{eqnarray}
   e_r^{(\pm)} 
    = N^{(\pm)} \left( \begin{array}{c}
                            \lambda|a|  \\
                            \kappa_{\pm} {a \over |a|}
                         \end{array}
                  \right)_r \,,
   \qquad \mbox{with} \qquad
   \kappa_\pm = \half(E_1-E_2) \pm \half\Delta E \,,
\label{eigenvec}
\end{eqnarray}
where the $N^{(\pm)}$ normalisation factor is chosen so that
$|e_1^{(\pm)}|^2 + |e_2^{(\pm)}|^2 = 1$. As 
$a = |a|\zeta_a$ ($\zeta_a = \pm 1, \pm i$) then as
expected any possible phase of the matrix element is contained in 
the eigenvectors, the energy must be real. Note that the components 
of the eigenvectors are related: $e_2^{(-)} = - e_1^{(+)}\zeta_a$ 
and $e_2^{(+)} = e_1^{(-)}\zeta_a$. We sketch their behaviour in 
Fig.~\ref{sketch_eigenvec_decay}.
Shown are sketches of eq.~(\ref{eigenvec}) for $e_1^{(-)\,2}$ and 
$e_2^{(-)\,2}$ against $p$ both for the free and interacting case
corresponding for the eigenvalues for $\Sigma \to N$ shown in
Fig.~\ref{sketch_decay}. While in the free case the components of
$\vec{e}^{(\pm)}$ remain constant (left panel) for the interacting
case (right panel) they flip as the momentum $p$ changes.


\section{Incorporating the spin index}


We now make a few comments on how the spin index is incorporated into
the formalism. Further details are given in \cite{qcdsf_fh}.
We have the replacement
\begin{eqnarray}
   |B_r(\vec{p_r})\rangle \to |B_r(\vec{p}_r,\sigma_r)\rangle \,,
\end{eqnarray}
where $\sigma_r = \pm$ is the spin index. Hence the $D$ matrix is
doubled in size $\sigma_r r = +1, -1, \ldots +d_S, -d_S$
i.e.\ we now have a $2d_S\times 2d_S$ matrix. However due to Kramers'
degeneracy theorem the energy states corresponding to 
$|B_r(\vec{p_r},\sigma_r)\rangle$, are doubly degenerate,
so we still have $d_S$ energy eigenvalues. 
We could continue as before with this enlarged matrix. However it is
advantageous to try to keep as close as possible to the previous results.
We can achieve this by writing the overlaps as
\begin{eqnarray}
   {}_\lambda\langle 0 | \hat{B}_{r\,\alpha}(\vec{0}) 
             | B_r(\vec{p}_r,\sigma_r) \rangle_{\lambda}
     &=& Z_r \, u^{(r)}_\alpha(\vec{p}_r,\sigma_r) + \ldots \,,
                                                         \nonumber  \\
   {}_\lambda\langle B_s(\vec{p}_s,\sigma_s) 
             | \hat{\bar{B}}_{s\,\beta}(\vec{0}) |0 \rangle_{\lambda}
     &=& \bar{Z}_s \, \bar{u}_\beta^{(s)}(\vec{p}_s,\sigma_s) + \ldots \,,
\end{eqnarray}
where $Z_r$ and $\bar{Z}_s$ are taken as scalars. Although the states 
here are the perturbed states, rather than the
unperturbed states, we expect the effect of the perturbation to be small
as from eq.~(\ref{state_lambda}) the $O(\lambda)$ terms involve overlaps
such as $\langle 0 |\hat{B}_r|Y\rangle$ or $\langle X|\hat{B}_r|B_r\rangle$ 
which vanish or are small due to the orthogonality of the spectrum.
Furthermore, although we could consider the Dirac indices as a GEVP 
it is convenient to sum over them with some matrix, $\Gamma$. 
Presently we only numerically consider the unpolarised case with
$\Gamma^{\rm unpol} = (1+\gamma_4)/2$ so
\begin{eqnarray}
   C_{\lambda\,rs}(t) = {\rm tr}\,\Gamma^{\rm unpol}C_{\lambda\,B_rB_s}(t) \,.
\end{eqnarray}
This reduces $D$ to the previous $d_S\times d_S$ matrix as in 
eq.~(\ref{D_def}) and leads to the replacement in eq.~(\ref{matrix_a})
of $a \to (a_{++}+a_{--})/2$ where the indices are the spin components.
So effectively this is the same result as before, we are just averaging 
over the spins. So this gives finally
\begin{eqnarray}
   C_{\lambda\,rs}(t)
      = \sum_{i=1}^{d_S} w^{(i)}_r\bar{w}_s^{(i)} e^{-E_\lambda^{(i)}t} \,.
\label{C_rs_nodirac}
\end{eqnarray}
Alternatively an explicit form factor decomposition of the matrix elements
(for all possible $\gamma$ matrices) shows that different spin components 
of matrix elements are related to each other. The upshot is that for 
previous examples in section~\ref{examples} for $d_S = 2$ we first make the
replacement
\begin{eqnarray}
   a \to \left( \begin{array}{rr}
                   a_{++}    & a_{+-}  \\
                   a_{-+}    & a_{--}
                \end{array}
         \right) \,,
\end{eqnarray}
with $a_{--} = \eta a^*_{++}$, $a_{-+} = -\eta a_{+-}^*$
($\eta = \pm$ depending on the matrix element considered)
and in the previous results we replace
$|a| \to |\det a|^{1/2}$ where $|\det a| = |a_{++}|^2 + |a_{+-}|^2$. 
As we pick out either $a_{++}$ or $a_{+-}$ this is equivalent to the
previous procedure.


\section{A lattice application for transition matrix elements}
\label{lattice}


As an example of this formalism, we shall now consider in more detail 
how the previous results can be applied to the $\Sigma \to N$ 
transition matrix element. We first discuss the necessary modifications 
to the action and the fermion inversion procedure before considering 
the specific numerical results.

To apply the results of section~\ref{examples} we need to consider the action
\begin{eqnarray}
   S = S_g + \int_x \left(\bar{u}, \bar{s}\right) 
                               \left( \begin{array}{cc}
                                         D_u         & -\lambda\cal{T} \\
                                         -\lambda\cal{T}^\prime
                                                        & D_s          \\
                                      \end{array}
                               \right) \left( \begin{array}{c}
                                                 u  \\
                                                 s  \\
                                              \end{array}
                                       \right) + \int_x \bar{d}\,D_d\,d \,,
\label{lat_trans_act}
\end{eqnarray}
where $S_g$ is the gluon action and the fermionic piece is explicitly
given. (For simplicity we absorb any clover terms into the $D$s.) 
We take the $u$ and $d$ quarks as mass degenerate $m_u = m_d \equiv m_l$, 
with a common mass $m_l$. For ${\cal T}$ we take the general local expression
\begin{eqnarray}
   {\cal T}(x,y;\vec{q}) = \gamma \, e^{i\vec{q}\cdot\vec{x}} \, \delta_{x,y} \,,
\end{eqnarray}
and for $\gamma_5$-hermiticity for the matrix in eq.~(\ref{lat_trans_act}) 
we need ${\cal T}^\prime = \gamma_5 {\cal T}^\dagger \gamma_5$. 

From the action in eq.~(\ref{lat_trans_act}) we see that we now need 
to invert a larger matrix to find the propagator for the various 
correlation functions. Although possible directly, we have found
it advantageous to consider it as a $2\times 2$ block matrix and invert that. 
This leads to 
\begin{eqnarray}
   G^{(uu)} &=& (1 - \lambda^2  D_u^{-1}{\cal T}
                    D_s^{-1}\gamma_5{\cal T}^\dagger\gamma_5)^{-1} D_u^{-1} \,,
                                                          \nonumber  \\
   G^{(ss)} &=& (1- \lambda^2 D_s^{-1}\gamma_5{\cal T}^\dagger\gamma_5 D_u^{-1} 
                             {\cal T})^{-1} D_s^{-1} \,,
\label{Guu+Gss_res}
\end{eqnarray}
and
\begin{eqnarray}
   G^{(us)} &=& \lambda D_u^{-1}{\cal T} G^{(ss)} \,,
                                                          \nonumber  \\
   G^{(su)} &=& \lambda D_s^{-1}\gamma_5{\cal T}^\dagger\gamma_5 G^{(uu)} \,.
\label{Gus+Gsu_res}
\end{eqnarray}
The problem with eq.~(\ref{Guu+Gss_res}) is that it involves an inversion
within an inversion, which computationally would be very expensive.
However for $\lambda$ small (the case considered here) it is sufficient
to expand to a low order in $\lambda$, especially as the expansion
parameter is $\lambda^2$. To build the Green's functions we use
$\delta_{\vec{x},\vec{0}}\delta_{t,0}$ as the initial source, and build the chain
using the previously calculated object as the new source. This has the 
advantage of producing the Green's function and hence correlation function
matrix
\begin{eqnarray}
   C_{\lambda\,rs}(t)
      = \left( \begin{array}{cc}
                  C_{\lambda\,\Sigma\Sigma}(t) &
                  C_{\lambda\,\Sigma N}(t)    \\
                  C_{\lambda\,N\Sigma}(t)     &
                  C_{\lambda\,NN}(t) 
               \end{array}
        \right)_{rs} \,,
\label{correl_fun_decay_mat}
\end{eqnarray}
as a continuous function of $\lambda$ rather than needing a separate 
evaluation for each value of $\lambda$ chosen. 

We now apply the GEVP (Generalised EigenValue Problem) to the $2 \times 2$
correlator matrix $C_\lambda(t)$, eq.~(\ref{correl_fun_decay_mat}). 
The variation of the method we use here, \cite{Owen:2012ts}, is first 
to determine the left $\varv^{(i)}$ and right $u^{(i)}$ eigenvectors by 
considering the correlation matrix at times $t_0$ and $t_0 + \Delta t_0$.
These can be combined with the correlator matrix
to construct a new correlation function
\begin{eqnarray}
   C_\lambda^{(i)}(t) = \varv^{(i)\,\dagger} C_\lambda(t)u^{(i)} \,, \quad i = \pm \,.
\end{eqnarray}
These two correlators $C_\lambda^{(i)}(t)$, $i =\pm$ represent the two GEVP
energy eigenstates of the system $\propto e^{-E^{(i)}_\lambda t}$ which of course 
includes the perturbation to the action. To relate this to the transition 
form factors, we require the energy splitting between these two states
and so from eqs.~(\ref{E_pm}), (\ref{DeltaE}) we construct the ratio 
of the correlators
\begin{eqnarray}
   R_\lambda(t)
      = { C_\lambda^{(-)}(t) \over C_\lambda^{(+)}(t) } 
      \,\, \stackrel{t\gg 0}{\propto} \,\, e^{-\Delta E_\lambda t} \,,
\label{ratio_R}
\end{eqnarray}
which in the large Euclidean time limit will behave like a 
single-exponential function and  will show up in the effective energy 
as a plateau region. We thus use this effective energy to pick out 
a suitable plateau region and then fit a single-exponential function 
to the ratio. The two important parameters of the GEVP calculation are 
$t_0$ and $\Delta t_0$. Optimally the time range from $t_0$ 
and $t_0+\Delta t_0$ needs to be in a region where the ground state is 
saturated but the signal-to-noise ratio is still sufficiently high 
to exclude any effects from higher states. Finally we note that
using eqs.~(\ref{w_simple}), (\ref{C_rs_nodirac}) means that
\begin{eqnarray}
   \varv_r^{(i)} = {N^{(i)} \over Z_r}\,e^{(i)}_r \,, \quad \mbox{and} \quad
   u_s^{(i)} = {\bar{N}^{(i)} \over \bar{Z}_s}\,e^{(i)}_s \,,
\label{u_v_e}
\end{eqnarray}
where $N^{(i)}$ and $\bar{N}^{(i)}$ are normalisation constants.
Essentially $\varv_r^{(i)*}$ measures the component of $B_r$ in the 
$i^{\rm th}$ eigenvector and similarly for $u_s^{(i)}$ and $\bar{B}_s$.


\section{Lattice results}


While the above discussion is general, we now consider the concrete case of
the vector matrix element $V_4$ for $\Sigma \to N$ where the $\Sigma$ 
is stationary, i.e.\ $\vec{p}_1 = \vec{0}$ and $\vec{p}_2 = \vec{q}$
in eq.~(\ref{transition_me}). Then the (Euclidean) momentum transfer 
is given in this case by%
\footnote{Note that we have adopted the convention that $q$ is
positive for a scattering process where for the scattered baryon the 
momentum $q$ is added to the initial baryon momentum. This is opposite 
to the semi-leptonic case, where the lepton and neutrino carry
momentum $q$.}
\begin{eqnarray}
   q = (i(M_\Sigma - E_N(\vec{q})), \vec{q}) \,,
   \quad \mbox{or} \quad
   Q^2 = - (M_\Sigma - E_N(\vec{q}))^2 + \vec{q}^2 \,.
\label{q_Q2_def}
\end{eqnarray}
Thus from eq.~(\ref{DeltaE}) we must compute
\begin{eqnarray}
   \Delta E_\lambda
     = \sqrt{ (E_{N} - M_{\Sigma})^2
               + 4\lambda^2 
                 \left( { \langle N(\vec{q})|\bar{u}\gamma_4 s
                                         |\Sigma(\vec{0})\rangle
                          \over
                          (2E_N)(2M_\Sigma) }^2 \right) } \,.
\label{DeltaE_decay}
\end{eqnarray}
Numerical simulations have been performed using $N_f = 2+1$ $O(a)$
improved clover Wilson fermions \cite{Cundy:2009yy} at $\beta = 5.50$ and
$(\kappa_l, \kappa_s) = (0.121040, 0.120620)$ on
a $N_s^3\times N_t = 32^3\times 64$ lattice. More definitions and 
details are given in \cite{Bietenholz:2011qq}. We just mention here 
that our strategy is to keep the average bare quark mass constant
from the $SU(3)$ flavour symmetric point. 
This situation corresponds to a lattice spacing of 
$a \sim 0.074\,\mbox{fm} \sim 1/(2.67\,\mbox{GeV})$ leading to a pion
mass of $\sim 330\,\mbox{MeV}$. Errors given in the following are 
primarily statistical (using $\sim O(500)$ configurations) using a 
bootstrap method.

Clearly we need to keep the energy states close to each other.
As spatial momentum on the lattice is discretised and given in each 
direction in steps of $2\pi/N_s$, which is coarse on this lattice size.
To obtain a finer energy level separation we use twisted boundary 
conditions, \cite{Bedaque:2004kc,Flynn:2007ess}, in the $y$-direction and set 
$\vec{q} = (0, \theta_2/N_s,0)$ and take $6$ values of the twist parameter
$\theta_2$ such that in lattice units $\vec{q}^2$ runs from $0$ (run \#1) 
to $\sim 0.05$ (run \#6), so that $Q^2 \sim -0.01\,\mbox{GeV}^2$ to 
$\sim 0.35\,\mbox{GeV}^2$.

\begin{figure}[htbp]
  \centering
  \begin{subfigure}[b]{0.45\textwidth}
    \centering
    \includegraphics[width=\textwidth]{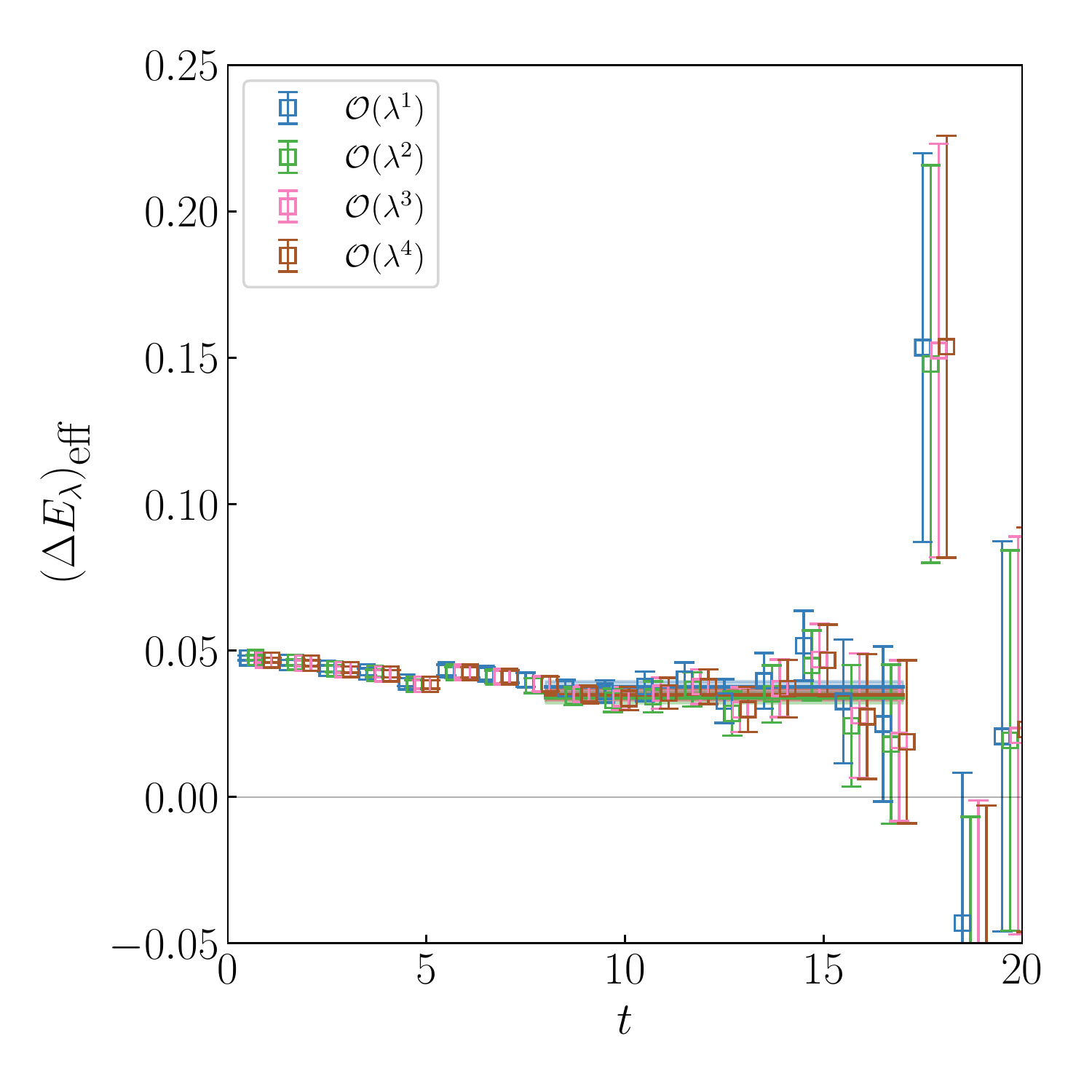}
  \end{subfigure}
  \begin{subfigure}[b]{0.45\textwidth}
    \centering
    \includegraphics[width=\textwidth]{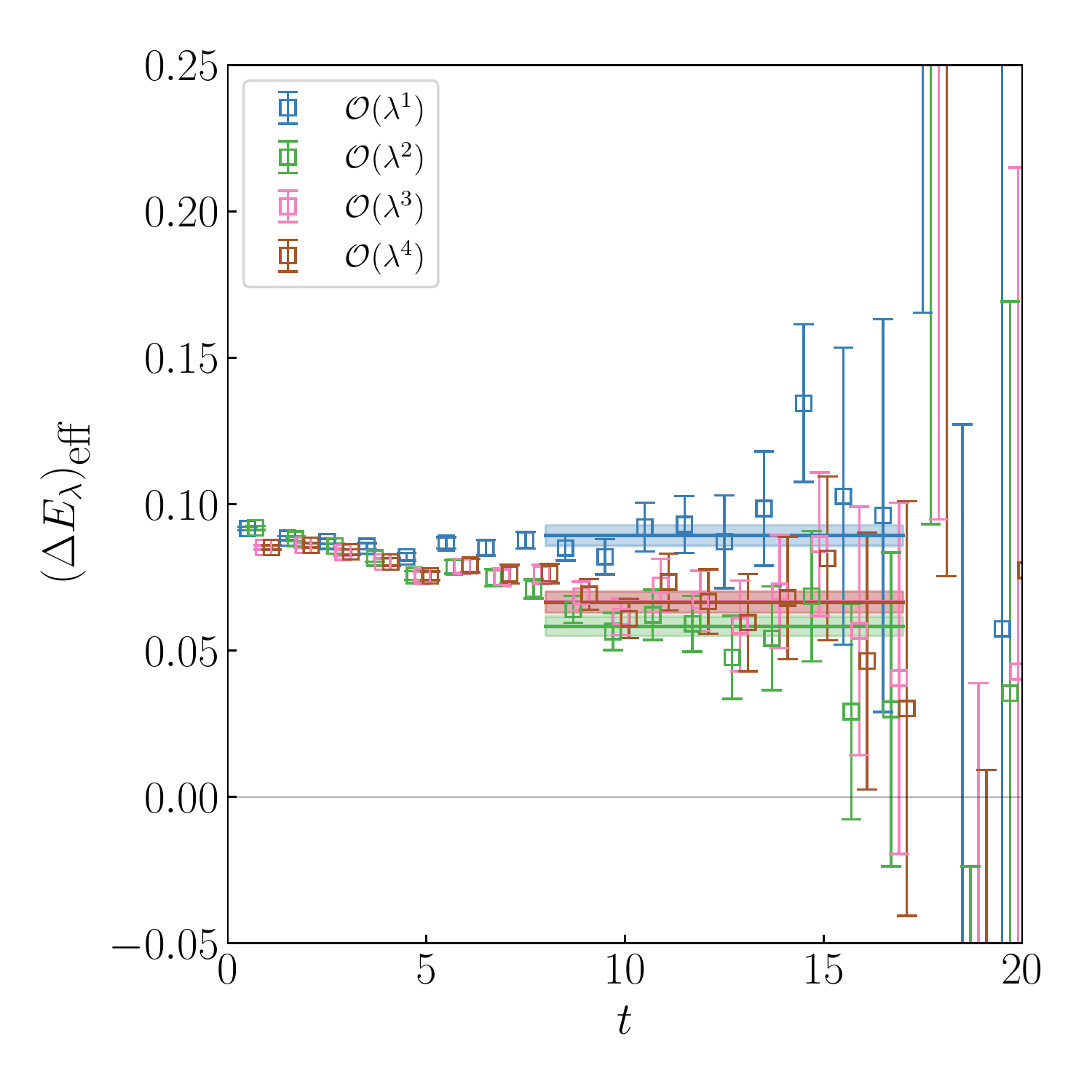}
  \end{subfigure}
  \vspace*{-0.125in}
  \caption{LH panel: 
           $(\Delta E_{\lambda})_{\rm eff} = -\ln(R_\lambda(t+1)/R_\lambda(t))$ 
           versus $t$ for $\lambda=0.025$ at $O(\lambda)$, $O(\lambda^2)$, 
           $O(\lambda^3)$ and $O(\lambda^4)$ for run \#5. 
           RH panel: similarly for $\lambda=0.05$. The points are 
           slightly offset for visibility.}
\label{fig:diffG-run5}
\end{figure}

Using each of these momentum values we calculate the correlation 
function matrix in eq.~(\ref{correl_fun_decay_mat}) up to order 
$O(\lambda^4)$ in the expansion of eqs.~(\ref{Guu+Gss_res}) and
(\ref{Gus+Gsu_res}). Since the multiplication with $\lambda$ occurs 
after the fermion matrix inversions, we are able to construct the 
correlation function matrix for a large number of $\lambda$ values 
in the range $\lambda = (0,\ldots,0.05)$. After solving the GEVP for each 
of these matrices we construct the ratio in eq.~(\ref{ratio_R}).
The effective energy of this ratio is shown in Fig.~\ref{fig:diffG-run5} 
for run \#5 at two different $\lambda$ values.
The right hand plot in this figure also shows the effect of the higher 
order corrections at $\lambda = 0.05$. 

\begin{figure}[htbp]
  \centering
  \begin{subfigure}[b]{0.45\textwidth}
    \centering
    \includegraphics[width=\textwidth]{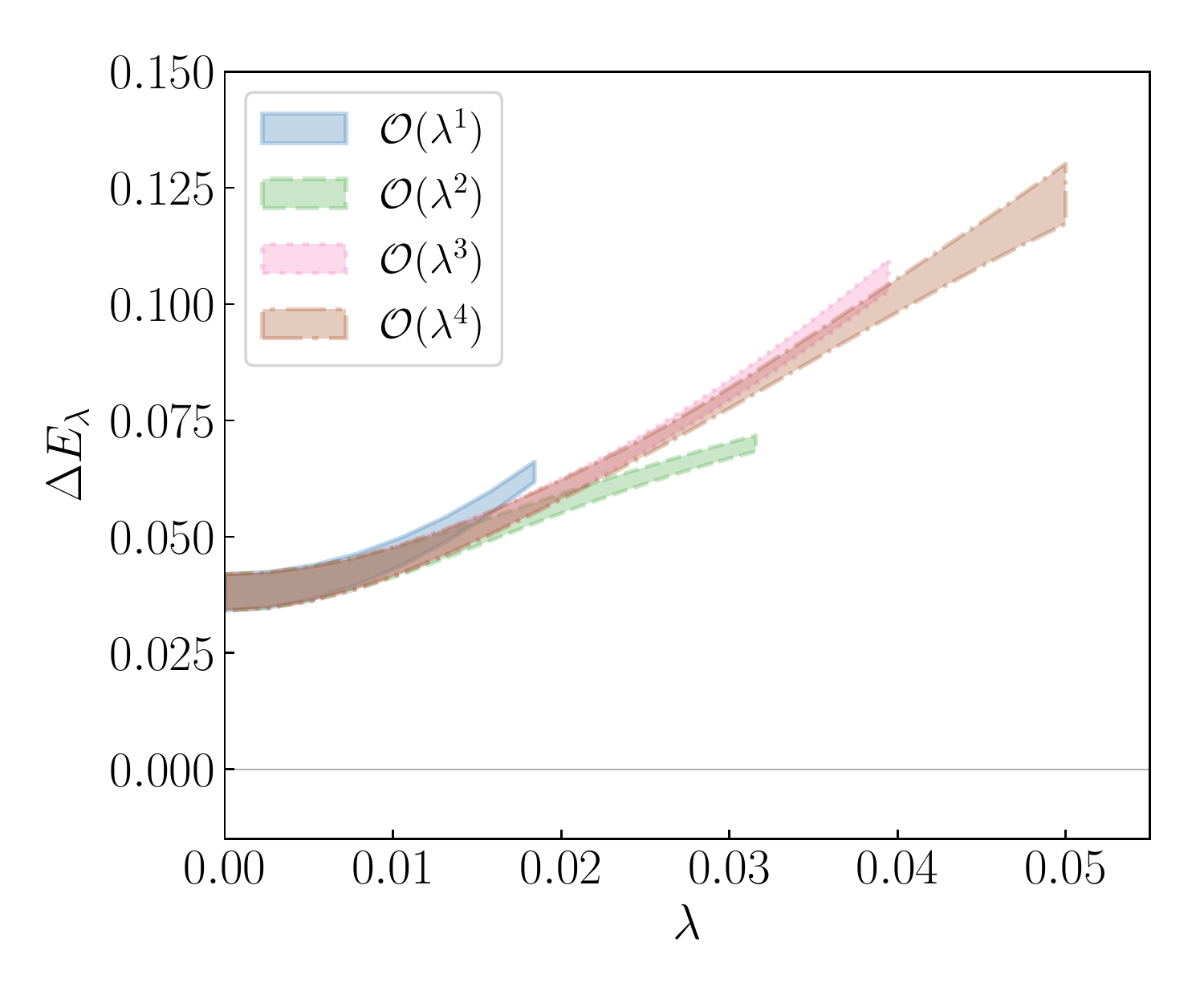}
  \end{subfigure}
  \begin{subfigure}[b]{0.45\textwidth}
    \centering
    \includegraphics[width=\textwidth]{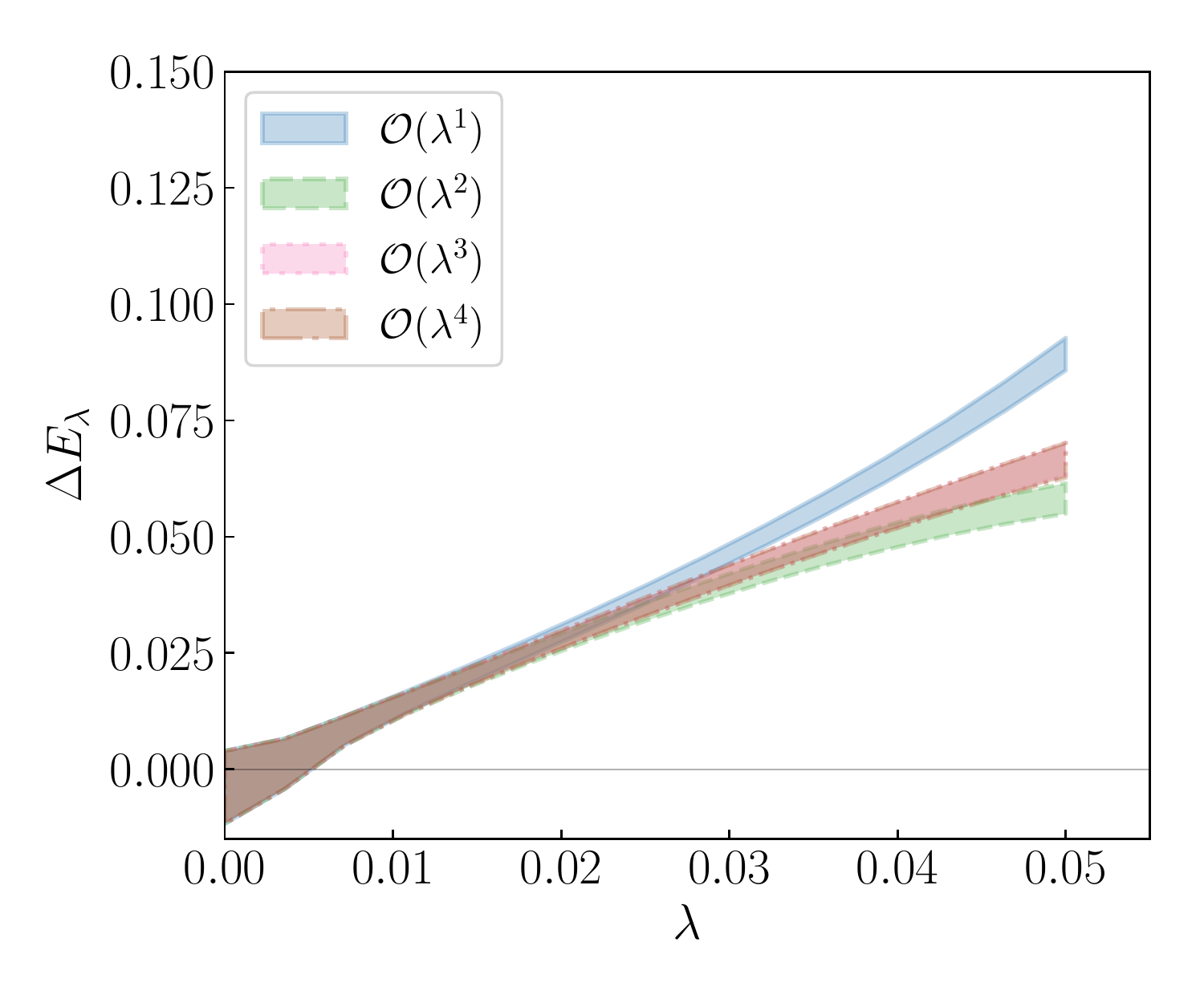}
  \end{subfigure}
  \caption{LH panel: The $\lambda$-dependence for run $\#1$ for 
           $\Delta E_\lambda$. The numerical results for each
           order in $\lambda$ ($O(\lambda)$, $O(\lambda^2)$, $O(\lambda^3)$
           and $O(\lambda^4)$) are given as bands.
           RH panel: Similarly for run $\#5$.}
\label{fig:lambda-dep}
\end{figure}

Figure \ref{fig:lambda-dep} shows the dependence of the energy shift 
on $\lambda$ for each of the four orders in the expansion for run \#1 
and run \#5. Once again we can see that as $\lambda$ increases the lower 
orders of the expansion start to deviate and higher order corrections 
are required. The expansion seems to hold up better for run \#5 where 
the energy gap between the unperturbed states is minimized. However 
even for run \#1, there is a sufficiently large $\lambda$ range available 
to extract the matrix element. 

\begin{figure}[htbp]
  \centering
  \includegraphics[width=0.55\textwidth]
                     {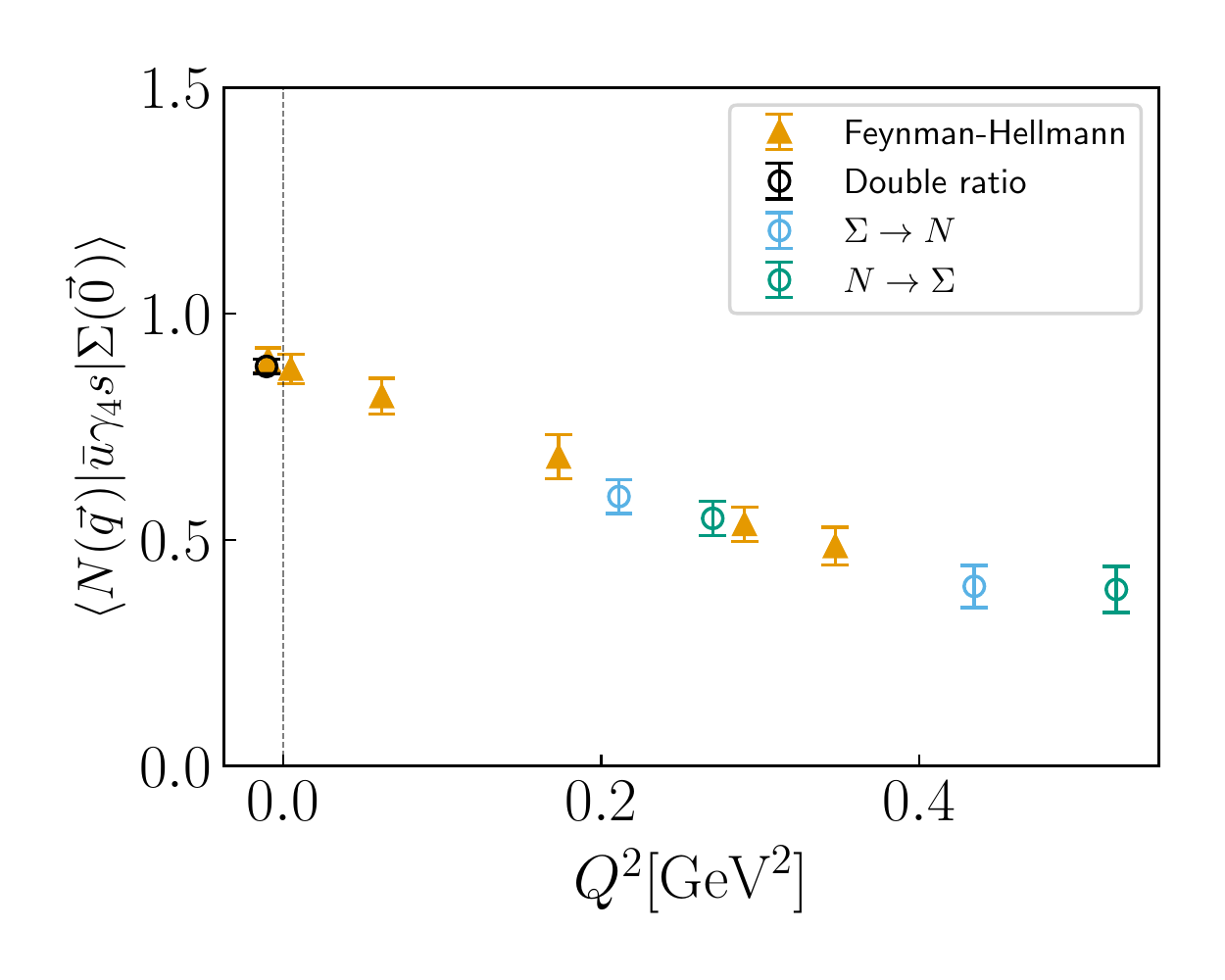}
  \vspace*{-0.125in}
  \caption{The renormalised transition matrix element as a function of 
           $Q^{2}$ from the Feynman-Hellmann method (triangles) and from 
           the three-point function method (circles).
           ($Z_V$ is taken from \protect\cite{Bickerton:2019nyz}.)}
\label{fig:matrix-element} 
\end{figure}

The matrix element can then be extracted by using eq.~(\ref{DeltaE_decay}),
the result of which is shown in Fig.~\ref{fig:matrix-element}. 
We also show the results of a three-point function calculation on the 
same configurations, there is good agreement between the two methods.


\section{Conclusions}


The Feynman-Hellmann approach has been shown here to be a viable alternative 
to the conventional three-point function method for calculating matrix elements.
The Feynman-Hellmann approach allows for a simpler analysis of excited 
state contributions as the resulting correlator has the same structure
as a two-point function. This allows for the application of the many 
established techniques for analysing two-point correlation functions. 
To extend this method to transition matrix elements has required 
reformulating it for quasi-degenerate states and using partially twisted 
boundary conditions to achieve these quasi-degeneracies.
The extention also allows for the inclusion of higher orders in the 
$\lambda$ expansion which has allowed us to extend the range of $\lambda$
which can be used. We have shown that this method can produce results with 
good agreement to the three-point function method for the $\Sigma\to N$ 
transition for $Q^{2}$ values  $-0.01\ \textrm{GeV}^{2}$ to 
$ 0.35\ \textrm{GeV}^{2}$.
Further details are given in \cite{qcdsf_fh}.


\section*{Acknowledgements}


The numerical configuration generation (using the BQCD lattice QCD 
program~\cite{Haar:2017ubh})) and data analysis (using the Chroma software 
library~\cite{Edwards:2004sx}) was carried out on the DiRAC Blue Gene Q 
and Extreme Scaling (EPCC, Edinburgh, UK) and Data Intensive (Cambridge, UK)
services, the GCS supercomputers JUQUEEN and JUWELS (NIC, J\"ulich, Germany)
and resources provided by HLRN (The North-German Supercomputer Alliance), 
the NCI National Facility in Canberra, Australia (supported by the 
Australian Commonwealth Government) and the Phoenix HPC service 
(University of Adelaide). RH is supported by STFC through grant ST/P000630/1.
HP is supported by DFG Grant No. PE 2792/2-1. PELR is supported in part 
by the STFC under contract ST/G00062X/1. GS is supported by 
DFG Grant No. SCHI 179/8-1. RDY and JMZ are supported by the 
Australian Research Council grant DP190100297.
For the purpose of open access, the authors have applied a Creative Commons 
Attribution (CC BY) licence to any author accepted manuscript version arising 
from this submission.



\end{document}